**Scalable *in operando* strain tuning in nanophotonic waveguides enabling three-quantum dot superradiance**

Joel Q. Grim[1*], Allan S. Bracker[1], Maxim Zalalutdinov[1], Samuel G. Carter[1], Alexander C. Kozen[2], Mijin Kim[3], Chul Soo Kim[1], Jerome T. Mlack[4], Michael Yakes[1], Bumsu Lee[4], Daniel Gammon[1]

[1] U.S. Naval Research Laboratory, Washington D.C.
[2] ASEE postdoctoral research fellow at the U.S. Naval Research Lab
[3] KeyW corporation
[4] NRC research associate at the U.S. Naval Research Lab

**ABSTRACT**

The quest for an integrated quantum optics platform has motivated the field of semiconductor quantum dot research for two decades. Demonstrations of quantum light sources, single photon switches, transistors, and spin-photon interfaces have become very advanced. Yet the fundamental problem that every quantum dot is different prevents integration and scaling beyond a few quantum dots. Here, we address this challenge by patterning strain via local phase transitions to selectively tune individual quantum dots that are embedded in a photonic architecture. The patterning is implemented with *in operando* laser crystallization of a thin $HfO_2$ film "sheath" on the surface of a GaAs waveguide. Using this approach, we tune InAs quantum dot emission energies over the full inhomogeneous distribution with a step size down to the homogeneous linewidth and a spatial resolution better than 1 μm. Using these capabilities, we tune multiple quantum dots into resonance within the same waveguide and demonstrate a quantum interaction via superradiant emission from three quantum dots.

**INTRODUCTION**

After much effort over the last twenty years, remarkable progress has been made in all aspects of using self-assembled quantum dots (QDs) for quantum information technologies, except one: scaling up the number of quantum dots in an integrated system. The combination of bandgap engineering with molecular beam epitaxy and modern nanofabrication techniques provides all the tools for building quantum optical systems on a chip. QDs are readily incorporated into solid state diodes for single electron injection[1] and into photonic architectures with waveguides, high Q cavities and superconducting detectors[2–4]. QDs can be deterministically positioned[5–7]. They have narrow linewidths and large optical dipole moments, couple strongly to solid state cavities[8,9], and are efficient deterministic sources of high quality single photons[10], entangled photons, and cluster states.[2,11–13] They have a spin memory[14] that interfaces with single photon emission or absorption, with their strong optical nonlinear response enabling single photon switching and control[15,16]. Entanglement between a spin qubit and a photon qubit[17–19], and between two distant spin qubits has been demonstrated[20,21].

And yet all this effort has remained focused on single QDs, or at most, two-QD systems. The reason for this is that the inhomogeneous linewidth is typically tens of meV, while the homogeneous linewidth, which determines how closely spectral matching must be done, is a few μeV. There is little hope of reducing the inhomogeneous linewidth sufficiently through improvements in growth, and the odds of finding a set of QDs with the same energy becomes negligible beyond a few QDs. Therefore, tuning of each individual QD is required. This is extremely challenging, not just because of the tuning range and resolution required, but also because it needs to be done within a photonic architecture.

Requirements for a scalable solution to QD spectral inhomogeneity include 1) energy tuning over the full inhomogeneous linewidth, 2) tuning resolution down to the homogeneous linewidth, 3) local tuning with a spatial resolution on the size scale of photonic features (1 µm), 4) an approach that does not degrade optical properties such as the QD emission linewidth, and 5) compatibility with photonic architectures.

Much attention has been devoted to overcoming the spectral inhomogeneity of QDs, yielding a variety of approaches that address a number of the requirements listed above. Short-range tuning has been accomplished using the Stark effect via electrical bias,[22] temperature,[23,24] AC Stark shift,[25] and Raman emission.[26,27] One of the most promising approaches is the application of strain using piezo-electric actuators that has enabled wide control of QD emission frequencies with high precision.[28–31] This comes at the cost of a relatively large fabrication overhead, especially with regard to tuning QDs independently on the same chip, with little prospect for independent tuning within the same waveguide or cavity. Dielectric capping layers have also been used to shift QD emission frequencies via strain, with thicker layers producing larger shifts,[32,33] but this approach does not permit real-time feedback or localized tuning. Rapid thermal annealing also provides long-range tuning by changing the composition of the QD.[34,35] Laser processing offers the prospect of local thermal annealing,[36,37] but the requisite high temperatures (> 650 °C) can damage the semiconductor material, leading to degraded QD emission. A promising laser-processing approach involves local laser-induced volume expansion of a phase change material, achieving reversable, moderate QD energy shifts (up to 2 meV) on bulk samples.[38]

Here, we demonstrate a laser-heating approach that satisfies all of the above requirements simultaneously. Using laser-induced modification of the crystal structure of a thin encapsulating layer of $HfO_2$, we introduce a controlled mechanical strain in specific areas of photonic structures with submicron spatial resolution. This is done to provide each QD of interest with a local strain field optimized to bring its emission energy to pre-selected values. We use InAs QDs embedded in GaAs bridge (Fig. 1) and photonic crystal (Fig. 4) waveguides that include n-i-n-i-p diodes, where the QD charge state can be deterministically controlled. By tuning the trion ($X^-$) optical transition of three QDs to the same emission energy within a single waveguide, we show that this approach is scalable, with the potential for creating quantum networks of many QDs with identical emission energies.

We demonstrate the first steps toward these networks by measuring superradiant coherent emission from three single-photon QD emitters coupled to the same waveguide and which have been tuned into resonance. Superradiance was first discovered theoretically in 1954 for dense clouds of atoms[39] and has been studied in great detail.[40] However, only recently has it become feasible to measure the superradiance of a small number of atoms.[41] Atoms emit indistinguishable photons, a requirement for superradiance, but are very difficult to localize. In contrast, QD emitters are relatively easy to localize in cavities and waveguides, but because they all have different emission frequencies, they are manifestly distinguishable. Recently, superradiance of two defects in an optical cavity in diamond and also two QDs in a waveguide have been demonstrated.[42,23] Our approach to emission energy tuning makes it straightforward to begin scaling up the number of emitters. Here, we show that it enables three-emitter superradiance, a milestone for solid-state systems.

**APPROACH**

We implement the strain-based tuning by conformally coating a suspended GaAs waveguide containing a low density of InAs QDs (~1/μm$^2$) with a thin and uniform amorphous film of HfO$_2$ via atomic layer deposition (ALD), see Fig.1. By tightly focusing a 532 nm cw laser on a region of the waveguide containing the target QD, the temperature is locally raised to the HfO$_2$ crystallization temperature (~400 °C) inducing crystallization within a region that can be smaller than the laser beam diameter. A complete amorphous to crystalline transition results in a roughly 3% reduction of HfO$_2$ volume[43] compressing the semiconductor within the partially crystallized sheath. This induced strain blue-shifts the QD emission energy with no measurable change in linewidth (see Supplement Fig. S5) and little effect on nearby QDs. The magnitude of the strain is defined by geometrical factors, the HfO$_2$ Young's modulus, and by the degree of crystallinity within laser-treated regions of the HfO$_2$ film. The latter can be controlled by adjusting the laser power and the exposure time.

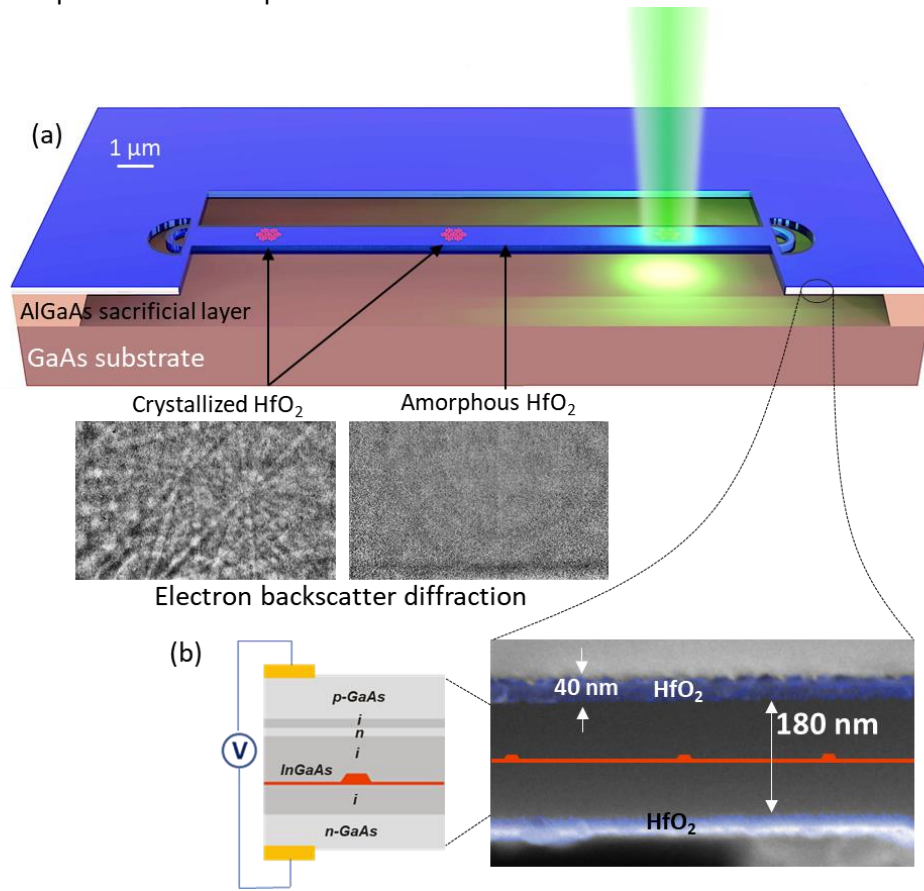

**Fig. 1 Schematic of the energy tuning approach and sample overview**. **a,** A laser is used to locally heat a GaAs bridge waveguide, which crystallizes a small region of an ALD HfO$_2$ thin film. The resulting compressive strain shifts the emission energy of a QD underneath the crystallized region. Electron backscatter diffraction measurements on and away from positions that have been laser heated confirm selective HfO$_2$ crystallization. **b,** Right: Cross-sectional SEM showing conformal ~40 nm HfO$_2$ film. Left: GaAs membrane heterostructure with n-i-n-i-p regions to deterministically charge the QDs with electrons.

After laser heating, electron backscatter diffraction provides clear signatures for crystallized HfO$_2$, indicated by the presence of Kikuchi lines for the laser-heated spot (Fig. 1a). Local crystallization is demonstrated by scanning the electron beam along the length of the waveguide, with a clear distinction between the amorphous and crystallized regions. The laser-heated spots were found to be polycrystalline from the random rotation of the Kikuchi line patterns for different positions within laser-heated regions (see Supplement Fig. S3). This implies crystallization from multiple nucleation sites, consistent with the crystallization behavior of ALD $HfO_2$ from previous studies[44]. In that scenario, the gradual increase in strain required to controllably shift the emission energy of a QD originates from increasing the volume fraction of crystallized $HfO_2$ in a given area. Electron diffraction measurements were also performed on other samples after rapid thermal annealing, from which a similar $HfO_2$ polycrystalline structure and an approximate crystallization temperature of 400°C were determined.

We find that film morphology is important for tuning behavior. Films deposited at 300°C provided energy tuning over a wider range, as well as better control than one deposited at 200°C. This can be understood from the morphology of the film that develops during deposition. The films deposited at 300°C were significantly rougher than those at 200°C as measured by scanning electron microscopy (SEM) and atomic force microscopy (see Supplement Fig. S1-2). An SEM image of the membrane surrounded by a ~40 nm $HfO_2$ film deposited with ALD at 300°C is shown in Fig. 1b. The rough surface structure may indicate that a dense network of crystal nucleation sites form during deposition at a temperature of 300°C.[44] In Ref. 44 a similar surface morphology corresponded to only a small volume fraction of crystalline $HfO_2$. From the rough structure measured with SEM and AFM, we hypothesize that there is an initial nucleation of crystallites during deposition[44] at 300°C of approximately 100/µm$^2$, which then readily grow in size upon laser heating to 400°C. For the 200°C film, nucleation is very sparse (see Supplement Fig. S1-2). The relatively fine-grained polycrystalline structure obtained at the higher deposition temperature is the likely source of the good resolution in the control of the QDs, both spatially and in energy that we will show below.

**QD ENERGY TUNING**

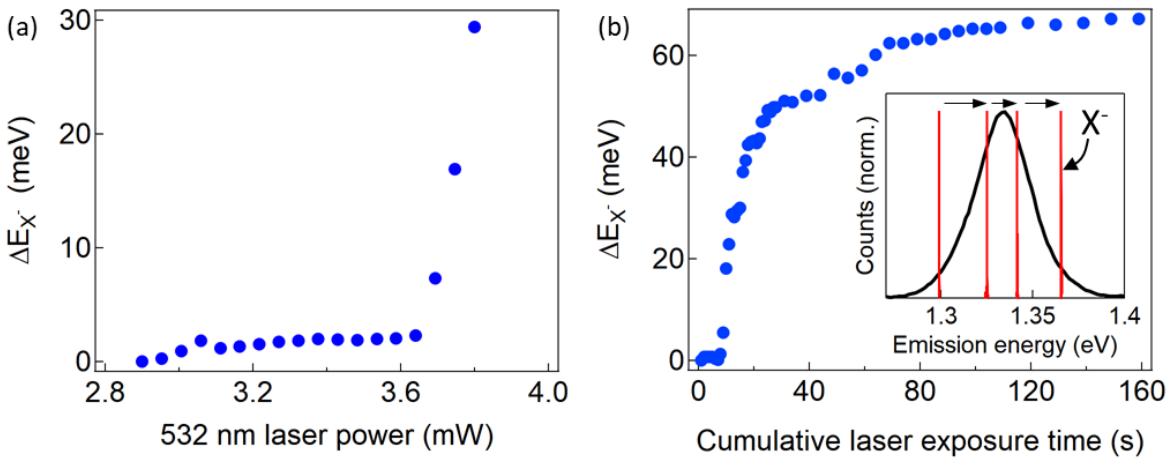

**Fig. 2 QD tuning curves. a,** Tuning the emission energy of a single QD in a 1.5 µm waveguide by increasing the heating laser power (the power was measured before the focusing objective), with the emission measured between heating laser exposures. It should be noted that the requisite laser power depends on the position on the waveguide, as well as the waveguide width. **b,** Tuning the emission energy of a single QD in a 1 µm waveguide via repeated exposures of a 532 nm laser (2.7-3 mW, 1-10s exposures). Points correspond to the change in X$^-$ emission energy measured between exposures. The inset shows the X$^-$ spectrum (red) at various points along the tuning curve compared to a typical ensemble spectrum (black) with a fwhm of 30 meV.

Because only a very small volume of the structure is heated, laser-induced crystallization can be initiated at cryogenic temperatures (~6 K), laser heated to 100s of °C, and cooled back down very quickly. Using this approach, the photoluminescence (PL) spectrum of an individual QD was probed with a second laser (890 nm) immediately after each heating exposure, with the QD already cooled back to low temperatures, providing real-time feedback. This approach is demonstrated in Fig. 2a by ramping the power of a 532 nm heating laser with 5s exposures, measuring the change in the $X^-$ emission energy between exposures. QD tuning curves obtained in this way serve as a calibration for the appropriate power to use for an optimized tuning procedure. Incrementally increasing the power even for very short exposure times leads to a highly nonlinear increase in emission energy with large steps and eventually the sudden destruction of the diode and membrane. Instead we find that better control and larger range can be achieved by multiple exposures at fairly low power (e.g. for the conditions in Fig. 2a, a power lower than the kink in the curve). With this optimization, it is possible to tune the emission of a QD more than 65 meV with small intermediate steps, as shown in Fig. 2b. The points in this figure are the $X^-$ emission energy peaks measured between exposures of the 532 nm heating laser (power: 2.7-3 mW) with a duration adjusted between 1-10 seconds. The inset in Fig. 2b compares the $X^-$ spectrum at various points along that tuning curve with a typical ensemble spectrum (fwhm ~30 meV), clearly demonstrating that the entire inhomogeneous distribution can be covered. We note that we also obtain significant energy shifts with thinner $HfO_2$ films. For example, even with a 5 nm thick film we obtain shifts of 1 meV. The conditions have not yet been optimized for thinner films and we focus here on results from 40 nm thick films for bridge waveguides. For photonic crystal waveguides, we use 10 nm thick $HfO_2$ films deposited at 300°C, with a tuning range of 5 meV. We find no discernable degradation of photonic crystal waveguide properties or cavity $Q$ factors (see Methods), indicating $HfO_2$ films thicker than the 10 nm used in this initial study should be possible.

**SPATIAL RESOLUTION**

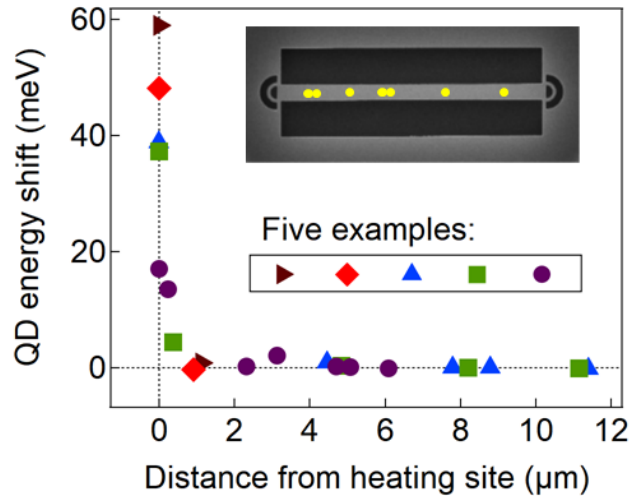

**Fig. 3 Spatial resolution of QD tuning**. Energy tuning as a function of distance from the laser-heated site showing the spatial resolution of the technique. The relative position of each QD is determined by fitting to the intensity of the laser in optical images of the waveguide (see Supplement Fig. S8), with five different examples shown. The yellow points shown on the inset SEM image mark the positions for the example shown with purple circles in this figure (with positions relative to the 5th point from the left).

Fig. 3 demonstrates the spatial resolution of independently tuning QDs in the same waveguide. This was accomplished by recording the spectra of QDs located in different positions along a waveguide and then tuning the emission energy of a selected QD. The spectra of the other QDs were then re-measured. The points in Fig. 3 indicate the change in the "bystander" QD peak emission energies after tuning a target QD. The QD positions were identified in optical images of the waveguide using the location of the probe laser at maximum QD emission intensity (see Supplement Fig. S8). The position of 0 μm corresponds to the location of a QD under the center of the heating laser, and the relative positions of the other QDs were recorded. For example, the yellow points on the inset SEM correspond to QD positions for the data shown with purple circles. The left-most point corresponds to the position of a QD that was tuned by 17 meV, after which the spectral shifts of the other QDs were recorded and plotted against their relative positions. Generally, we find that there is very little shift in QDs that are more than 1 μm from the center of the heating laser spot. In fact, we have cases in which a second QD is within the laser spot and shifts very little. For example, the first two green square points in Fig. 3 show that the energy shift (37.25 meV) for a QD at the center of the heating laser spot is 8× greater than a QD < 400 nm away (4.3 meV). Therefore, we conclude that under appropriate conditions the spatial resolution can be 1 μm or less. We find that the spatial resolution for photonic crystal waveguides is similar. Finite element modeling of the thermal profile and the mechanical strain associated with $HfO_2$ crystallization provides additional insight into these results and is shown in Fig. S9 in the Supplement.

**TUNING QDS INTO RESONANCE**

We have tuned QDs into resonance within both bridge waveguides (Fig. S10) and photonic crystal waveguides (PhC WGs). QDs at different positions were excited nonresonantly by a NIR laser that was coupled into the waveguide via a grating coupler (Fig. 4a). The QD PL was collected from a coupler at the other end of the waveguide. A 532 nm laser was focused onto the position of a chosen QD along the waveguide, with the power and exposure time adjusted to crystallize $HfO_2$ and fine-tune the emission energy of the QD into resonance with other QDs in the same waveguide. The emission was sent through a high-resolution fiber Fabry-Perot interferometer during tuning to resolve the spectral alignment of the QDs. Tuning QDs into resonance with each other within a single waveguide requires not only sufficient tuning range and high spatial resolution, but also fine spectral tuning resolution. Step sizes of 1 μeV with a standard deviation of about 1 μeV are currently possible as shown in the Supplement (Fig. S11). We are currently limited by the stability of our system, and to a lesser extent, by film morphology.

We demonstrate tuning multiple QDs into resonance in Fig. 4. In Fig. 4b,c, the $X^-$ emission energies of two QDs were initially 0.54 meV out of resonance (> 60 times the 8.7 μeV linewidth of QD1), as shown in the high-resolution Fabry-Perot interferometer spectrum in Fig. 4b. QD1 was strain-tuned into resonance with QD2 at a fixed bias of 0.5V, with the top three spectra in Fig. 4c showing three tuning steps, with both QDs resonant in the third spectrum (middle, black circles). Red and blue lines in the top two spectra are fits to the QD1 and QD2 emission peaks, respectively. We spectrally detune the two QDs via the Stark effect, as shown in the bottom two spectra in Fig. 4c, providing a convenient reversible way to probe the transition from indistinguishable to distinguishable photons. The bias-dependent linewidth-broadening observed for QD1 can be attributed to co-tunneling near the charge stability edge.

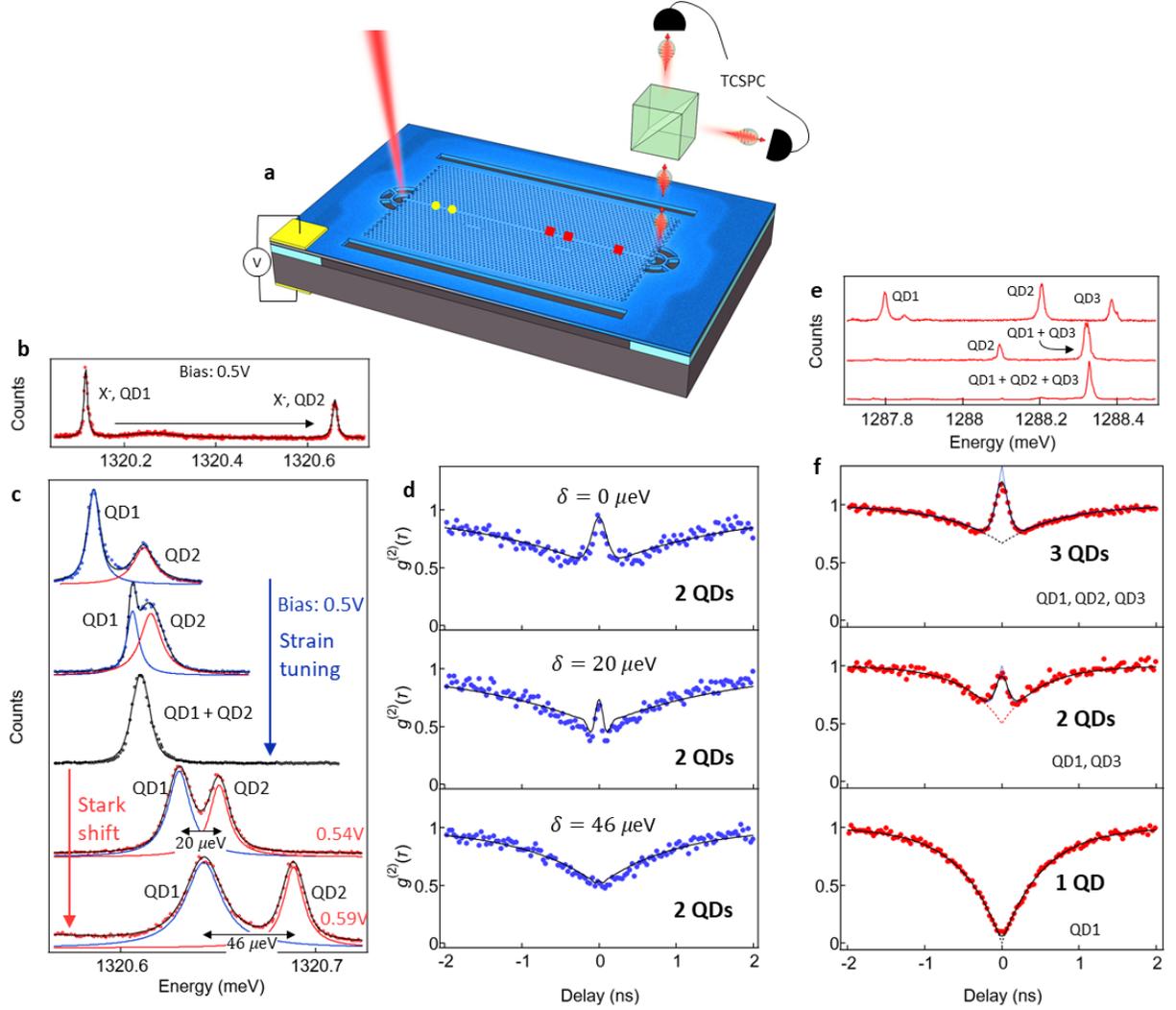

**Fig. 4 Superradiant emission from strain-tuned QDs in a photonic crystal waveguide. a,** SEM image of a PhC waveguide and schematic with the positions of the QDs indicated with yellow circles (corresponding to QDs in b, c, d), and red squares (corresponding to QDs in a different waveguide for e and f). **b,** High-resolution Fabry-Perot interferometer spectra of the X$^-$ charge state for two QDs located 1.2 μm apart prior to strain-tuning. **c,** Spectra of the same QDs in the final stages of strain tuning (top three traces), with the final peak comprising both QDs in resonance. The Stark-shifted spectra (bottom two curves) were obtained by tuning the electric field, providing a convenient way to distinguish the QDs once they are in resonance. **d,** Second-order correlation measurements for resonant and detuned cases (corresponding to the bottom three spectra in **c**). **e,** Fabry-Perot interferometer spectrum showing the X$^-$ charge state for three QDs (in a different PhC waveguide than example shown in **b**-**d**). The red-shifted final spectrum with respect to the blue-most target QD is a consequence of a small initial shift of all QDs in the WG that occurred after the first heating laser exposure. **f,** Second-order photon correlation measurements for three (top), two (middle), and one (bottom) QDs.

**SUPERRADIANCE IN A PHOTONIC CRYSTAL WAVEGUIDE**

As a first example of the power of our tuning technique, we demonstrate superradiance of up to three QDs into a photonic crystal waveguide mode. After tuning two QDs into resonance using strain (Fig. 4c, top to middle), a second-order correlation measurement ($g^{(2)}$) was carried out (Fig. 4d) and then repeated for the two bias-controlled detunings (δ) shown in bottom of Fig. 4c. The measurement essentially gives the normalized counts on one detector at time $\tau$ after the detection of a count on the other. For two distinguishable single photon emitters, $g^{(2)}$ will have an *antibunching* dip, going to 0.5 at $\tau = 0$. For two resonant emitters ($\delta = 0$ µeV), there is a sharp *bunching* peak around $\tau = 0$, which is a clear signature of superradiance.

In Fig. 4f, we take this a step further by displaying $g^{(2)}(\tau)$ for a system of *three* QDs in a PhC WG at different steps in the tuning process (emission spectra are shown in Fig. 4e). First, one QD is measured alone, while detuned from the other two QDs, showing a typical antibunching curve going down nearly to zero. Then, two of the QDs are tuned into resonance and measured, showing a reduction in the antibunching dip, and a clear central peak. Finally, after all three QDs are tuned into resonance, the antibunching dip becomes quite shallow, and the central peak gets stronger. This behavior is quite robust, having been observed in all attempts (five pairs of QDs and two trios in different WGs).

A calculation of $g^{(2)}(\tau)$ for $N$ emitters coupled to the same waveguide mode[23,45] in the bad cavity limit gives:

$$g^{(2)}(\tau) = \left[1 - \frac{1}{N}e^{-\gamma\tau}\right] + \frac{1}{N^2}e^{-2\Gamma\tau - 4\pi^2\sigma^2\tau^2}\sum_{i \neq j}\cos(\delta_{ij}\tau) \qquad (1)$$

where the radiative emission rate is $\gamma$, while $\Gamma = \gamma/2 + \gamma_{pd}$ is the contribution to the linewidth due to the radiative rate and pure dephasing, and $\sigma$ is the contribution to the linewidth from spectral diffusion[46]. The detuning between pairs of emitters is $\delta_{ij}$. Eq. (1) is a simplified model in which each QD is assumed to have the same radiative rate, linewidth, excitation amplitude and coupling to the waveguide. A more general expression is discussed in the Methods.

The first term in Eq. 1 describes the independent contribution from each of the QDs as if they were distinguishable, resulting in an antibunching dip. The dip at $\tau = 0$ increases as $(1 - 1/N)$ with the number of QDs. The second term gives the interference between QDs, producing the coherent bunching peak superimposed on the antibunching dip. This sharp bunching peak in $g^{(2)}$ is a result of the coherent interaction of the resonant emitters through the emission process. For example, when two indistinguishable single-photon emitters are both excited, the emission of a single photon into the waveguide mode can come from either emitter, and the two are left in an entangled state with a coherence time determined by the spectral linewidths. During this coherence time, the emission of a second photon can come immediately with an emission rate twice that of distinguishable emitters. This leads to the sharp coherent peak in $g^{(2)}$. From Eq. 1, the bunching peak should rise to $g^{(2)}(0) = 2(1 - 1/N)$, giving $g^{(2)}_{N=2}(0) = 1$ and $g^{(2)}_{N=3}(0) = 4/3$, signatures of superradiance that are clearly evident in our results shown in Fig. 4d,f.

In Fig. 4d, we experimentally explore the transition between indistinguishable to distinguishable QD emission by measuring $g^{(2)}(\tau)$ for detunings $\delta = 0, 20$, and 46 µeV between two QDs. When $\delta = 0$, the interference term leads to a sharp bunching peak as shown in the data and the corresponding fit to Eq. 1, with $g^{(2)}(0) = 0.95$ very close to the ideal value of $g^{(2)}(0) = 1$ for two indistinguishable QDs. The difference can be accounted for by the 100 ps time resolution of our measuring system. The width of the measured coherent bunching peak (200 ps) is determined by decoherence, which is also manifested in the linewidths of the QDs. Significant detuning of the QDs leads to fast oscillations in the second term, which are largely averaged out by the finite time resolution of the measurement system. As a result of these fast oscillations for $\delta = 46$ µeV, the coherent bunching peak is averaged out by the finite time resolution and the $g^{(2)}$ data shows the characteristic antibunching behavior of two distinguishable emitters, with $g^{(2)}(0) = 0.5$. For a modest detuning of $\delta = 20$ µeV, the coherent bunching peak is sharper and weaker, consistent with averaging over oscillations. Excellent agreement with Eq. 1 (solid lines) is obtained with $\sigma = 1$ ns$^{-1}$, $\gamma_{pd} = 2.5$ ns$^{-1}$ and $\gamma = 0.7$ ns$^{-1}$ for $\delta = 0, 20$ µeV and $1$ ns$^{-1}$ for $\delta = 46$ µeV in Fig. 4d.

For the system of three QDs, the measured $g^{(2)}(\tau)$ is also in excellent agreement with the model for superradiance, as shown in Fig. 4f. For all three QDs in resonance, Eq. 1 gives $g^{(2)}(0) = 4/3$. The measured coherent bunching peak is $g^{(2)}(0) = 1.2$, with the difference from the model consistent with the measurement time resolution, as shown with the light blue line. By setting the coherent term to zero in the fit, we obtain the distinguishable contribution to $g^{(2)}(\tau)$ as shown by the dotted curves in Fig. 4f, with $g^{(2)}(0)$ very close to the ideal values of 2/3 (three QDs) and 1/2 (two QDs). Again, the width of the coherent peak is determined by the linewidth, agreeing very well with Eq. 1 using the same values of $\sigma = 1$ ns$^{-1}$ and $\gamma_{pd} = 2.5$ ns$^{-1}$ for all fits, and $\gamma = 1.9, 2,$ and $1.4$ ns$^{-1}$ for the one, two, and three QD data, respectively.

The spectral diffusion contribution to the linewidth is currently the most significant limitation affecting the indistinguishability of the QDs. In the above fits, the spectral diffusion contribution was 10 µeV FWHM which is consistent with the average spectral linewidths of the QDs. We find that under resonant excitation the spectral linewidths decrease by about a factor of three in this waveguide sample. Therefore, measurements under resonant excitation will enhance the indistinguishability and presumably broaden the width of the coherent bunching peak. Moreover, resonant excitation would enable single-photon Raman emission, which would allow additional spectral fine-tuning as well as single-photon pulse shaping.[47]

**DISCUSSION AND OUTLOOK**

We have demonstrated a scalable approach to overcome the spectral inhomogeneity of InAs QDs, which has been one of the most difficult challenges not only for QDs but all solid-state quantum emitters. The approach involves patterning strain to locally tune QD emission energies via laser-induced crystallization of a thin film deposited on the surface of a photonic structure. We have shown that QDs can be tuned across the inhomogeneous distribution with a resolution down to the homogenous linewidth (see Supplement Fig. S11). Combined with the sharp spatial resolution, we have shown that multiple QDs can be tuned into spectral alignment within a single waveguide, enabling the measurement of superradiance of three quantum dots.

The principles of this approach are general and can be extended to other photonic structures such as nanowires[32,33] and micropillars[48]. There is also the prospect to address spectral inhomogeneity in other emitter classes, for example defects in diamond[49]. The strain sensitivity of the optical transitions of the NV center in diamond[50], for example, is similar to QDs used in this work[51], but has an inhomogeneous distribution much narrower than QDs, which could compensate for the higher Young's modulus in diamond.

We have chosen to demonstrate this technique with HfO$_2$ and have demonstrated that its properties are well suited to this application. The crystallization temperature (~400°C) is fairly low, and well below the melting temperature of GaAs and also where there is significant migration of beryllium and silicon in our diode structure. Furthermore, with a Young's modulus almost twice that of GaAs, much of the strain is taken up in the GaAs, while the ability to conformally coat around the membrane with ALD enhances the effect. Nevertheless, there may be other materials that are as good or better, such as materials with a lower crystallization temperature or reversible tuning.[38,52]

We anticipate that this work will substantially impact the development of scalable InAs QD-based photon sources, and could enable on-chip photonic quantum information processing.[53] The deterministic charging enabled by the diode structure in our samples allowed us to demonstrate superradiance with the X$^-$ charge state. An important milestone that may now be within reach is an extended network of entangled electron (hole) spin qubits connected with on-chip flying photon qubits.

**METHODS**

**Sample structure**
InGaAs QDs were grown by molecular beam epitaxy within a membrane diode heterostructure on an n-doped GaAs substrate. The diode allows injection of electrons into the QD. The heterostructure consists of a 0.5 μm Si-doped (n-type) GaAs buffer, 950 nm of Si-doped $Al_{0.7}Ga_{0.3}As$, 30 nm Si-doped n-type GaAs; 30 nm undoped GaAs; InGaAs QDs grown with a partial cap (nominal 2.7 nm) followed by a brief anneal; 71 nm undoped GaAs; 10 nm n-type Si-doped GaAs; 10 nm undoped GaAs; and 30 nm p-type Be-doped GaAs. The intermediate n-type layer in this n-i-n-i-p diode reduces the forward bias required to charge the QD, avoiding high currents through the device[54,55] The waveguide patterns are produced using electron beam lithography and a $SiCl_4$-based inductively coupled plasma. The $Al_{0.7}Ga_{0.3}As$ is undercut with 10% hydrofluoric acid. Electrical contacts are made with indium to the p-type layer and to the back of the n-doped substrate. The waveguides are 184 nm thick, 15 μm long and either 1, 1.5 or 2 μm wide. They are terminated with semi-circular grating outcouplers.[51] The photonic crystal membranes consisted of a triangular lattice of 66 nm radii holes with a lattice constant of 244 nm. The 15 μm long waveguide consisted of a row of missing holes with grating couplers on both ends.

**$HfO_2$ ALD**
$HfO_2$ thin films were deposited after lithography, but before mounting samples on a chip carrier and making electrical contacts. Films were deposited using commercial ALD tool at both 200°C and 300°C by atomic layer deposition (ALD) using Tetrakis(ethylmethylamido)hafnium(IV) (TEMAH) and $H_2O$ as metalorganic and oxidation precursors. The TEMAH and $H_2O$ were maintained at temperatures of 75°C and 20°C, respectively. The $HfO_2$ growth rate was 0.8 Å/cycle, measured by spectroscopic ellipsometry using a Cauchy optical model fit from 193 nm - 1650 nm. Film thickness was measured on Si 110 monitor coupons included in the same deposition batches as the GaAs substrates, after the samples had cooled to room temperature in the chamber load lock but prior to air exposure.

**Measurement system and laser-induced crystallization of $HfO_2$**
Samples were measured in a closed-cycle cryostat operating at 6K using a confocal microscope with mutually-aligned 532 nm (spot size: ~ 1 μm) heating and 890 nm NIR probe paths. QD tuning was automated by shuttering the green heating laser path and measuring the NIR-excited photoluminescence bias map. Although the spectral resolution of this system is 50 μeV, the emission peak position can be determined with a resolution of 1 μeV. A fiber Fabry-Perot interferometer with a 2.4 μeV resolution was used for higher resolution measurements of QDs that had been tuned to the same energy. The spectra of QDs within a waveguide were simultaneously measured by exciting and collecting on output couplers. The position of the QDs in waveguides was determined with a resolution of <170 nm from the location of the probe laser in optical images.

The heating laser power required to achieve the appropriate temperature depends on the thermal transport properties of the photonic structure. For suspended waveguides, the center of the waveguide is farthest from the bulk GaAs substrate heat sink and therefore requires a factor of 1.5-2 lower power than near the edges. The tuning procedure was optimized by ramping the laser power on different QDs located in different waveguide positions with automated control of the laser power, exposure time, and measurement of NIR laser-excited QD photoluminescence bias maps. The QD energy shift vs. power for different waveguide positions (see, for example Fig. 2a) served as starting points for subsequent QD tuning, enabling more efficient long- or short-range tuning. With a 40 nm film, we find that we can consistently achieve shifts greater than 30 meV, with more than 20 examples thus far. The rate of the energy shift for a given laser power depends on a number of factors, including the position on the waveguide, variations in QD strain sensitivity, and the laser-processing history. Thinner $HfO_2$ films allowed smaller, but significant energy tuning ranges (20nm: 6 meV, 5 nm: 1 meV). In the photonic crystal sample, a 10 nm film allowed a 5 meV tuning range.

A liquid crystal noise eater was used to stabilize the green laser power because relatively small power fluctuations can significantly affect the local temperature. The QD energies were stable while the sample was maintained at low temperature for ~4 months. Thermally cycling the sample up to room temperature resulted in QD red shifts on the order of 150 µeV, including QDs that were not tuned. This may originate from strain built in by the sample mounting procedure in which silver epoxy is used to mount it to the chip carrier, and thus may be mitigated in the future.

The spatial extent of crystallization determined from electron diffraction measurements included the effect of electron beam drift due to GaAs charging. However, repeated scans over the same crystallized regions yielded consistent values of < 0.5 µm. In addition to the polycrystalline nature of a crystallized region evidenced by the Kikuchi patterns (see Supplement Fig. S4), it was also observed that there were positions within a crystallized region that did not display Kikuchi patterns. This supports the conclusion of crystal growth from multiple nucleation sites.

We have focused on tuning the PL transition energies, however, there are additional changes in the spectra. While there is no measurable change in the homogeneous linewidths, there are changes in the relative energies and onset voltages of transitions arising from different charge states in the PL bias maps that arise from strain-induced changes in confinement. These have been studied in detail in earlier work with piezo-electric actuators,[29,30] and we simply summarize these effects in the Supplement.

The photonic crystal waveguide samples had the same diode heterostructure and QD vertical position as the bridge waveguide samples. For these samples, 10 nm of $HfO_2$ was deposited with ALD with no discernable degradation of waveguide properties or cavity *Q* factors (4000-6000) compared to another piece of the same sample without $HfO_2$. Quantum interference measurements were performed by exciting QDs in the waveguide by coupling 1330-1393 meV CW laser light into a grating coupler, and sending the emission from the opposite coupler through a monochromator to a 50:50 beamsplitter and single photon detectors.

**Superradiance Model and Measurement**
In Eq. 1 all the QDs were treated as identical. A more general equation can be derived[45,23,46] which accounts for differences in radiative rates, linewidths, excitation rate and coupling to the waveguide.

$$g^{(2)}(\tau) = \left[1 - \frac{\sum_i I_i^2 e^{-\gamma_i \tau}}{(\sum_i I_i)^2}\right] + \frac{\sum_{i \neq j} I_i I_j \, e^{-i\Gamma_{ij}\tau - 2\pi^2 \sigma_{ij}^2 \tau^2} \cos(\delta_{ij})}{(\sum_i I_i)^2}$$

Here, $\gamma_i$ is the radiative rate of the $i$th QD, $\Gamma_{ij} = (\gamma_i + \gamma_j)/2 + (\gamma_{pdi} + \gamma_{pdj})$ and $\sigma_{ij}^2 = \sigma_i^2 + \sigma_j^2$. $I_i$ is the measured intensity of the $i$th QD, and is proportional to the product of the QD excitation rate and its coupling to the waveguide.

The first term in this expression models the independent contributions of the QDs, while the second arises from the coherent interference between each pair of QDs. The magnitudes of both terms at $\tau = 0$ decrease when the emission intensities are unequal, resulting in a larger antibunching dip and smaller bunching peak. All the QDs were excited by a single laser traveling through the waveguide. However, by scanning the laser frequency (1330 -1393 meV), we were able find an energy that roughly equalized the measured emission intensities for QDs in this sample (via the excitation rate). When all the $I_i$ are equal, they drop out of the equation and the maximum $g^{(2)}(0)$ is obtained. We note that $g^{(2)}$ does not depend on the positions of the QDs.

Because of the relatively large spectral diffusion contribution to the linewidth in comparison to the radiative rate, the coherent bunching peak is much narrower than the antibunching dip. As a result, these features are essentially fit independently. All the data in the two data sets in Figs. 5d and 5f were fit with the same values for $\sigma_{ij}^2 = 2\sigma^2$ and $\Gamma_{ij} = \gamma + 2\gamma_{pd}$, except that $\gamma_i = \gamma$ was allowed to vary for each $g^{(2)}$ curve, resulting in physically reasonable values for $\gamma$. The difference in $\gamma$ between Figs. 5d and 5f is consistent with the fact they were different waveguides and the QDs were tuned to different energies (see Fig. S12 in Supplement). Although the linewidth of each QD was different, the calculation is relatively insensitive to this, and a single average value close to the measured linewidth of the resonant QDs worked well. We took the spectral diffusion to be $\sigma = 1$ ns$^{-1}$, corresponding to a spectral linewidth[46] of FWHM= $2\sqrt{2\ln(2)}\,\sigma h$ = 10 µeV, and the pure dephasing $\gamma_{pd} = 2.5$ ns$^{-1}$, corresponding to a linewidth of FWHM = $2\hbar\gamma_{pd}$= 3 µeV. These values are consistent with the average measured linewidths.

The data that support the findings of this study are available from the corresponding author upon reasonable request.

**Code availability**

COMSOL code used to simulate thermal and strain profiles is available from the corresponding author upon request.

**Acknowledgements**
This work was supported by the U.S. Office of Naval Research and the OSD Quantum Sciences and Engineering Program. A.C.K. acknowledges the support of the American Society for Engineering Education and U.S. Naval Research Laboratory postdoctoral fellowship program. J.T.M. and B.L. acknowledge the support of the NRC Research Associateship Program at the U.S. Naval Research Laboratory.

**References**
1. Warburton, R. J. *et al.* Optical emission from a charge-tunable quantum ring. *Nature* **405,** 926–929 (2000).
2. Lodahl, P., Mahmoodian, S. & Stobbe, S. Interfacing single photons and single quantum dots with photonic nanostructures. *Rev. Mod. Phys.* **87,** 347–400 (2015).
3. Petruzzella, M. *et al.* Quantum photonic integrated circuits based on tunable dots and tunable cavities. *APL Photonics* **3,** 106103 (2018).
4. Liu, F. *et al.* High Purcell factor generation of indistinguishable on-chip single photons. *Nat. Nanotechnol.* **13,** 835–840 (2018).
5. Kiravittaya, S., Rastelli, A. & Schmidt, O. G. Advanced quantum dot configurations. *Reports Prog. Phys.* **72,** (2009).
6. Yakes, M. K. *et al.* Leveraging crystal anisotropy for deterministic growth of InAs quantum dots with narrow optical linewidths. *Nano Lett.* **13,** 4870 (2013).
7. Jöns, K. D. *et al.* Triggered indistinguishable single photons with narrow line widths from site-controlled quantum dots. *Nano Lett.* **13,** 126 (2013).
8. Yoshie, T. *et al.* Vacuum Rabi splitting with a single quantum dot in a photonic crystal nanocavity. *Nature* **432,** 200–203 (2004).
9. Reithmaier, J. P. *et al.* Strong coupling in a single quantum dot-semiconductor microcavity system. *Nature* **432,** 197 (2004).
10. Senellart, P., Solomon, G. & White, A. High-performance semiconductor quantum-dot single-photon sources. *Nature* **12,** 1026–1039 (2017).
11. Huber, D. *et al.* Strain-Tunable GaAs Quantum Dot: A Nearly Dephasing-Free Source of Entangled Photon Pairs on Demand. *Phys. Rev. Lett.* **121,** 33902 (2018).
12. Schwartz, I. *et al.* Deterministic generation of a cluster state of entangled photons. *Science,* **354,** 434–437 (2016).
13. Aharonovich, I., Englund, D. & Toth, M. Solid-state single-photon emitters. *Nat. Photonics* **10,** 631–641 (2016).
14. Kroutvar, M. *et al.* Optically programmable electron spin memory using semiconductor quantum dots. *Nature* **432,** 81–84 (2004).
15. Sun, S., Kim, H., Luo, Z., Solomon, G. S. & Waks, E. A single-photon switch and transistor enabled by a solid-state quantum memory. *Science,* **361,** 57–60 (2018).
16. Javadi, A. *et al.* Spin-photon interface and spin-controlled photon switching in a nanobeam waveguide. *Nat. Nanotechnol.* **13,** 398–403 (2018).
17. De Greve, K. *et al.* Quantum-dot spin-photon entanglement via frequency downconversion to telecom wavelength. *Nature* **491,** 421 (2012).

18. Gao, W. B., Fallahi, P., Togan, E., Miguel-Sanchez, J. & Imamoglu, A. Observation of entanglement between a quantum dot spin and a single photon. *Nature* **491,** 426 (2012).
19. Schaibley, J. R. *et al.* Demonstration of Quantum Entanglement between a Single Electron Spin Confined to an InAs Quantum Dot and a Photon. *Phys. Rev. Lett.* **110,** 167401 (2013).
20. Stockill, R. *et al.* Phase-Tuned Entangled State Generation between Distant Spin Qubits. *Phys. Rev. Lett.* **119,** 10503 (2017).
21. Delteil, A. *et al.* Generation of heralded entanglement between distant hole spins. *Nat. Phys.* **12,** 218–223 (2016).
22. Pagliano, F. *et al.* Dynamically controlling the emission of single excitons in photonic crystal cavities. *Nat. Commun.* **5,** 5786 (2014).
23. Kim, J. H., Aghaeimeibodi, S., Richardson, C. J. K., Leavitt, R. P. & Waks, E. Super-radiant emission from quantum dots in a nanophotonic waveguide. *Nano Lett.* **18,** 4734–4740 (2018).
24. Faraon, A. *et al.* Local quantum dot tuning on photonic crystal chips. *Appl. Phys. Lett.* **90,** 213110 (2007).
25. Muller, A., Fang, W., Lawall, J. & Solomon, G. S. Creating polarization-entangled photon pairs from a semiconductor quantum dot using the optical stark effect. *Phys. Rev. Lett.* **103,** 217402 (2009).
26. Sweeney, T. M. *et al.* Cavity-stimulated Raman emission from a single quantum dot spin. *Nat. Photonics* **8,** 442 (2014).
27. Fernandez, G., Volz, T., Desbuquois, R., Badolato, A. & Imamoglu, A. Optically tunable spontaneous Raman fluorescence from a single self-assembled InGaAs quantum dot. *Phys. Rev. Lett.* **103,** 1–4 (2009).
28. Yuan, X. *et al.* Uniaxial stress flips the natural quantization axis of a quantum dot for integrated quantum photonics. *Nat. Commun.* **9,** (2018).
29. Ding, F. *et al.* Tuning the exciton binding energies in single self-assembled InGaAs/GaAs quantum dots by piezoelectric-induced biaxial stress. *Phys. Rev. Lett.* **104,** 67405 (2010).
30. Kuklewicz, C. E., Malein, R. N. E., Petroff, P. M. & Gerardot, B. D. Electro-Elastic Tuning of Single Particles in Individual Self-Assembled Quantum Dots. *Nano Lett.* **12,** 3761–3765 (2012).
31. Höfer, B. *et al.* Independent tuning of excitonic emission energy and decay time in single semiconductor quantum dots. *Appl. Phys. Lett.* **110,** 151102 (2017).
32. Stepanov, P. *et al.* Large and Uniform Optical Emission Shifts in Quantum Dots Strained along Their Growth Axis. *Nano Lett.* **16,** 3215–3220 (2016).
33. Bouwes Bavinck, M. *et al.* Controlling a Nanowire Quantum Dot Band Gap Using a Straining Dielectric Envelope. *Nano Lett.* **12,** 6206–6211 (2012).
34. Ellis, D. J. P. *et al.* Control of fine-structure splitting of individual InAs quantum dots by rapid thermal annealing. *Appl. Phys. Lett.* **90,** 2005–2008 (2007).
35. Nakaoka, T. *et al.* Tuning of g -factor in self-assembled In ( Ga ) As quantum dots through strain engineering. *Phys. Rev. B* **71,** 205301 (2005).
36. Rastelli, A. *et al.* In situ laser microprocessing of single self-assembled quantum dots and optical microcavities. *Appl. Phys. Lett.* **90,** 73120 (2007).
37. Fiset-Cyr, A. *et al.* In-situ tuning of individual position-controlled nanowire quantum dots

via laser-induced intermixing. *Appl. Phys. Lett.* **113,** 53105 (2018).
38. Takahashi, M. *et al.* Local control of emission energy of semiconductor quantum dots using volume expansion of a phase-change material. *Appl. Phys. Lett.* **102,** 93120 (2013).
39. R.H.Dicke. Coherence in Spontaneous Radiation Process. *Phys. Rev.* **93,** 99–110 (1954).
40. Mcardell, B. W. *et al.* The Super of Superradiance. *Science (80-. ).* **325,** 1510–1512 (2009).
41. Bridge, N. Superradiance for Atoms Trapped along a Photonic Crystal Waveguide. *Phys. Rev. Lett.* **63601,** 1–5 (2015).
42. Sipahigil, A. *et al.* An Integrated Diamond Nanophotonics Platform for Quantum Optical Networks. *Science,* **354,** 847–850 (2016).
43. Venkatachalam, D. K., Bradby, J. E., Saleh, M. N., Ruffell, S. & Elliman, R. G. Nanomechanical properties of sputter-deposited $HfO_2$ and $Hf_xSi_{1-x}O_2$ thin films. *J. Appl. Phys.* **110,** 43527 (2011).
44. Hausmann, D. M. & Gordon, R. G. Surface morphology and crystallinity control in the atomic layer deposition (ALD) of hafnium and zirconium oxide thin films. *J. Cryst. Growth* **249,** 251–261 (2003).
45. Lounis, B., Orrit, M., Koganov, G. A. & Shuker, R. Few emitters in a cavity: from cooperative emission to individualization. *New J. Phys.* **13,** 93020 (2011).
46. Kambs, B. & Becher, C. Limitations on the indistinguishability of photons from remote solid state sources. *New J. Phys.* **20,** 115003 (2018).
47. Pursley, B. C., Carter, S. G., Yakes, M. K., Bracker, A. S. & Gammon, D. Picosecond pulse shaping of single photons using quantum dots. *Nat. Commun.* **9,** 115 (2018).
48. Somaschi, N. *et al.* Near-optimal single-photon sources in the solid state. *Nat. Photonics* **10,** 340–345 (2016).
49. Sohn, Y. I. *et al.* Controlling the coherence of a diamond spin qubit through its strain environment. *Nat. Commun.* **9,** 17–22 (2018).
50. Lee, K. W. *et al.* Strain Coupling of a Mechanical Resonator to a Single Quantum Emitter in Diamond. *Phys. Rev. Appl.* **6,** 1–17 (2016).
51. Carter, S. G. *et al.* Sensing flexural motion of a photonic crystal membrane with InGaAs quantum dots. *Appl. Phys. Lett.* **111,** (2017).
52. Wuttig, M., Bhaskaran, H. & Taubner, T. Phase-change materials for non-volatile photonic applications. *Nat. Photonics* **11,** 465–476 (2017).
53. O'Brien, J. L., Furusawa, A. & Vučković, J. Photonic quantum technologies. *Nat. Photonics* **3,** 687–695 (2009).
54. Vora, P. M. *et al.* Spin–cavity interactions between a quantum dot molecule and a photonic crystal cavity. *Nat. Commun.* **6,** 7665 (2015).
55. Löbl, M. C. *et al.* Narrow optical linewidths and spin pumping on charge-tunable close-to-surface self-assembled quantum dots in an ultrathin diode. *Phys. Rev. B* **96,** 1–10 (2017).